# NANOELECTRONIC DEVICES: A UNIFIED VIEW

Supriyo Datta
School of Electrical and Computer Engineering,
Purdue University, West Lafayette, IN 47907.

## ABSTRACT

Nanoscale electronic devices are of great interest for all kinds of applications like switching, energy conversion and sensing. The objective of this chapter, however, is not to discuss specific devices or applications. Rather it is to convey the conceptual framework that has emerged over the last twenty years, which is important not only because of the practical insights it provides into the design of nanoscale devices, but also because of the conceptual insights it affords regarding the meaning of resistance and the essence of all non-equilibrium phenomena in general. We present a unified description applicable to a wide variety of devices from molecular conductors to carbon nanotubes to silicon transistors covering different transport regimes from the ballistic to the diffusive limit, based on what we call the NEGF-Landauer approach.

## 1. INTRODUCTION

Since "everyone" has a computer these days and every computer has nearly a billion Field Effect Transistors (FET's) working in concert, it seems safe to say that the most common electronic device is an FET, which is basically a resistor consisting of an active region called the channel with two very conductive contacts at its two ends called the source and the drain (Fig.1). What makes it more than just a resistor is the fact that a fraction of a volt applied to a third terminal called the gate changes the resistance by several orders of magnitude. Electrical switches like this are at the heart of any computer and what has made computers more and more powerful each year is the increasing number of switches that have been packed into one by making each switch smaller and smaller. For example a typical FET today has a channel length (L) of ~ 50 nm, which amounts to a few hundred atoms!

Nanoscale electronic devices have not only enabled miniature switches for computers but are also of great interest for all kinds of applications including energy conversion and sensing. The objective of this chapter, however, is not to discuss specific devices or applications. Rather it is to convey the conceptual framework that has emerged over the last twenty years, which is important not only because of the practical insights it provides into the design of nanoscale devices, but also because of the conceptual insights it affords regarding the meaning of resistance and the essence of all non-equilibrium phenomena in general.

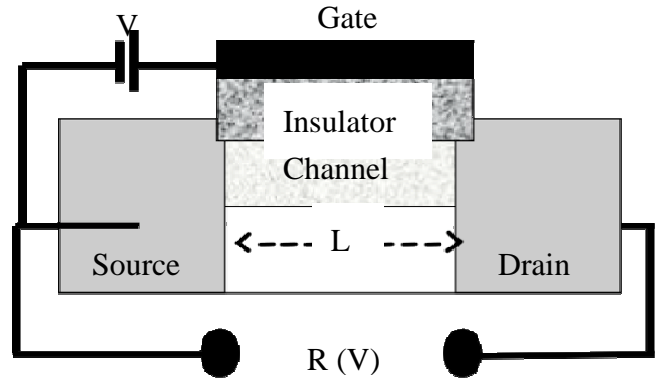

Fig.1: Schematic representing a Field Effect Transistor (FET), which consists of a channel with two contacts (labeled "source" and "drain"), whose resistance R can be controlled through a voltage V applied to a third terminal labeled the "gate", which ideally carries negligible current.

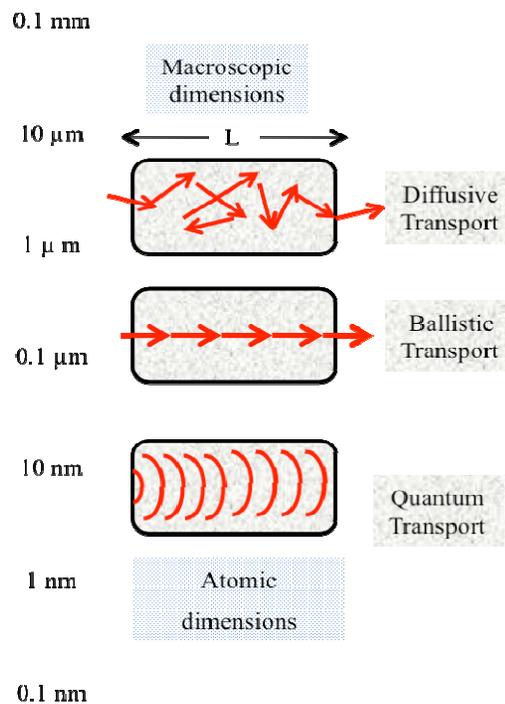

Fig.2: As the length "L" of the channel in Fig.1 is reduced the nature of electronic transport from one contact to the other changes qualitatively from diffusive to ballistic to quantum.

This new conceptual framework provides a unified description for all kinds of devices from molecular conductors to carbon nanotubes to silicon transistors covering different transport regimes from the diffusive to the ballistic limit (Fig.2). As the channel length L is reduced, the nature of electronic transport changes qualitatively. With long channels, transport is ***diffusive***, meaning that the electron gets from one contact to another via a random walk, but as the channel length is reduced below a mean free path, transport becomes ***ballistic***, or "bullet-like". At even shorter lengths the wave nature of electrons can lead to ***quantum*** effects such as interference and tunneling. Historically our understanding of electrical resistance and conduction has progressed **top-down**: from large macroscopic conductors to small atomic scale conductors. Indeed thirty years ago it was common to argue about what, if anything, the concept of resistance meant on an atomic scale. Since then there has been significant progress in our understanding, spurred by actual experimental measurements made possible by the technology of miniaturization. However, despite this progress in understanding the flow of current on an atomic scale, the standard approach to the problem of electrical conduction continues to be top-down rather than bottom-up. This makes the problem of nanoscale devices appear unduly complicated, as we have argued extensively [Datta 2005, 2008]. The purpose of this chapter is to summarize a unified **bottom-up** viewpoint to the subject of electrical conduction of particular relevance to nanoelectronic devices.

The viewpoint we wish to discuss is summarized in Fig.3a: Any nanoelectronic device has an active "channel" described by a Hamiltonian [H] which also includes any potential U due to other charges, external (on the electrodes) or internal (within the channel). The channel communicates with the source and drain (and any additional contacts) that are maintained in local equilibrium with specified electrochemical potentials. The communication between the channel and the contacts is described by the self-energy matrices $[\Sigma_1]$ and $[\Sigma_2]$ [Caroli et al. 1972]. Finally there is a self-energy matrix $[\Sigma_s]$ describing the interaction of an individual electron with its surroundings, which unlike $[\Sigma_{1,2}]$ has to be calculated self-consistently. Each of these quantities is a matrix whose dimension (NxN) depends on the number of basis functions (N) needed to represent the channel. How these matrices are written down varies widely from one material to another and from one approach (semi-empirical or first principles) to another. But once these matrices have been written down, the procedure for calculating the current and other quantities of interest is the same, and in this chapter we will stress this generic procedure along with its conceptual underpinnings.

The schematic model of Fig.3a includes both the diffusive (Fig.3b) and the ballistic (Figure 1.3c) limits as special cases. In the ballistic limit, the flow of electrons is controlled by the contact terms $[\Sigma_1]$ and $[\Sigma_2]$, while the interactions within the channel are negligible. By contrast, in the diffusive limit, the flow of electrons is controlled by the interactions within the channel described by $[\Sigma_s]$ and the role of contacts ($[\Sigma_1]$ and $[\Sigma_2]$) is negligible. Indeed prior to 1990, theorists seldom bothered even to draw the contacts. Note that there is an important distinction between the Hamiltonian matrix [H] and the self-energy matrices $[\Sigma_{1,2,s}]$. The former is Hermitian representing conservative dynamical forces, while the latter is non-Hermitian and helps account for the "entropic forces". Let me elaborate a little on what I mean by this term.

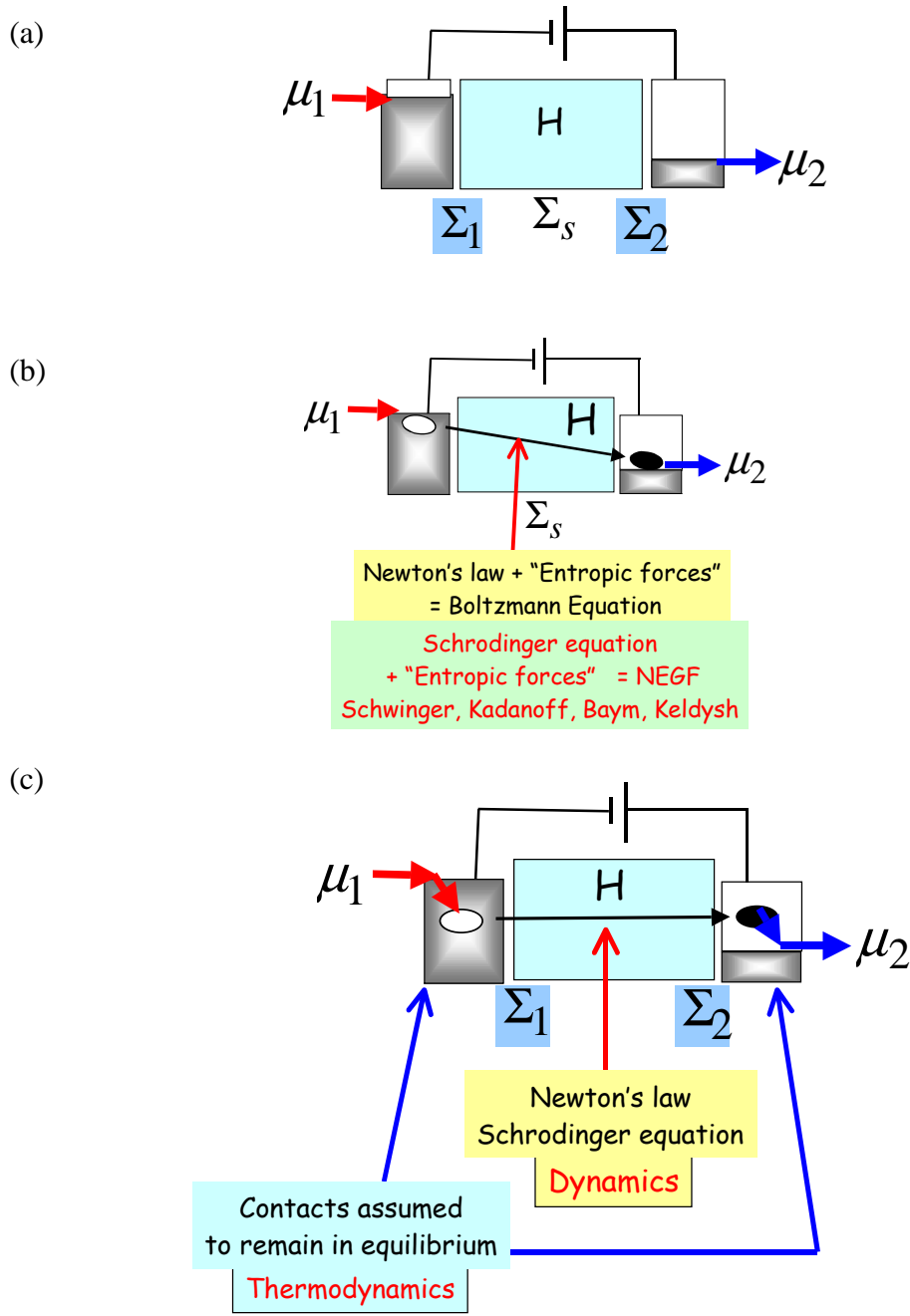

Fig.3(a): Schematic representing the general approach used to model nanoscale devices: The channel is described by a Hamiltonian [H] while the communication between the channel and the contacts is described by the self-energy matrices [$\Sigma_1$] and [$\Sigma_2$]. The self-energy matrix [$\Sigma_s$] describes the interaction of an individual electron with its surroundings. (b) In traditional long devices it is common to ignore the contacts, while (c) in the coherent limit a "Landauer model" neglecting incoherent interactions within the channel is more appropriate (Adapted from Datta 2005).

Consider a simple system like a hydrogen atom having two energy levels separated by an energy $\varepsilon_2 - \varepsilon_1$ that is much larger than the thermal energy $k_B T$ (Fig.4). We all know that an electron initially in the upper level $\varepsilon_2$ will lose energy, possibly by emitting a photon, and end up in the lower level $\varepsilon_1$, but an electron initially in the lower level $\varepsilon_1$ will stay there forever. Why? This tendency of all systems to relax unidirectionally to the lowest energy is considered so "obvious" that only a beginning student would even raise the question. But it is important to recognize that this property does not follow from the Schrodinger equation. Hamiltonians are always Hermitian with $|H_{12}| = |H_{21}|$. Any perturbation that takes a system from $\varepsilon_2$ to $\varepsilon_1$ will also take it from $\varepsilon_1$ to $\varepsilon_2$. The unidirectional transfer from $\varepsilon_2$ to $\varepsilon_1$ is the result of an entropic force that can be understood by noting that our system is in contact with a reservoir having an enormous density of states $D_r(E_r)$ that is a function of the reservoir energy $E_r$ [Feynman 1972]. Using E to denote the total energy of the reservoir plus the system, we can write the reservoir density of states as $D_r(E-\varepsilon_1)$ and $D_r(E-\varepsilon_2)$ corresponding to the system energy levels $\varepsilon_1$ and $\varepsilon_2$ respectively (Fig.4).

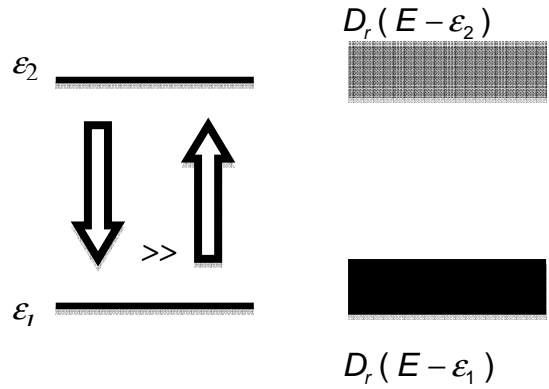

Fig.4: A system with two energy levels [$\varepsilon_1$] and [$\varepsilon_2$], coupled to a reservoir whose corresponding density of states are $D_r(E-\varepsilon_1)$ and $D_r(E-\varepsilon_2)$. The downward transition rate from $\varepsilon_2$ to $\varepsilon_1$ far exceeds the upward transition rate $\varepsilon_1$ to $\varepsilon_2$ although the Schrodinger equation would have predicted them to be equal. The unidirectionality arises from entropic forces as discussed in the text.

The ratio of the downward to the upward transition rate is given by

$$\frac{R_{2\to 1}}{R_{1\to 2}} = \frac{D_r(E-\varepsilon_1)}{D_r(E-\varepsilon_2)}$$

Why is the downward rate far greater that the upward rate: $R_{2\to1} >> R_{1\to2}$? Simply because for all normal reservoirs, the density of states is an increasing function of the reservoir energy so that with $(E-\varepsilon_1)>>(E-\varepsilon_2)$, we have $D_r(E-\varepsilon_1)>>D_r(E-\varepsilon_2)$. We call this an entropic force because the density of states is related to the entropy through the Boltzmann relation ($S = k_B \ln \Omega$):

$$\frac{D_r(E-\varepsilon_1)}{D_r(E-\varepsilon_2)} = \exp\left(\frac{S(E-\varepsilon_1) - S(E-\varepsilon_2)}{k_B}\right) \approx \exp\left(\frac{\varepsilon_2 - \varepsilon_1}{k_B} \frac{dS}{dE}\right)$$

Noting that the temperature T is defined as $1/T = dS/dE$, we can write

$$\frac{R_{2\to 1}}{R_{1\to 2}} = \frac{D_r(E-\varepsilon_1)}{D_r(E-\varepsilon_2)} = \exp\left(\frac{\varepsilon_2 - \varepsilon_1}{k_B T}\right) \qquad (1)$$

so that with $\varepsilon_2 - \varepsilon_1 \gg k_B T$, $R_{2\to 1} \gg R_{1\to 2}$ and the system relaxes to the lower energy as "everyone" knows.

The point I am trying to make is that the Schrodinger equation alone is not enough even to describe this elementary behavior that we take for granted. Like numerous other phenomena in everyday life, it is driven by entropic forces and not by mechanical forces. Clearly any description of electronic devices, quantum or classical, must incorporate such entropic forces into the dynamical equations. Over a century ago, Boltzmann showed how to combine entropic forces with Newton's law, and his celebrated equation still stands as the centerpiece in the transport theory of dilute gases, though it was highly controversial in its day and its physical basis still provokes considerable debate [see for example, McQuarrie 1976]. The non-equlibrium Green's function (NEGF) formalism, we describe in this chapter, originating in the work of Martin and Schwinger 1959, Kadanoff and Baym 1962 and Keldysh 1965, can be viewed as the quantum version of the Boltzmann equation: it combines entropic forces with Schrodinger dynamics.

What makes both the Boltzmann and the NEGF formalisms conceptually challenging is the intertwining of dynamical and entropic forces. By contrast, the ballistic limit leads to a relatively simple model with dynamical and entropic processes separated spatially. Electrons zip through from one contact to the other driven purely by dynamical forces. Inside the contacts they find themselves out of equilibrium and are quickly restored to equilibrium by entropic forces, which are easily accounted for simply by legislating that electrons in the contacts are always maintained in local equilibrium. We could call this the "Landauer model" after Rolf Landauer who had proposed it in 1957 as a conceptual tool for understanding the meaning of resistance, long before it was made experimentally relevant by the advent of nanodevices. Today there is indeed experimental evidence that ballistic resistors can withstand large currents because there is negligible Joule heating inside the channel. Instead the bulk of the heat appears in the contacts, which are large spatial regions capable of dissipating it. I consider this separation of the dynamics from the thermodynamics to be one of the primary reasons that makes a bottom-up viewpoint starting with ballistic devices pedagogically attractive.

Our objective is to present the complete NEGF-Landauer model for nanodevices (Figure 3a) that incorporates the contacts into the classic NEGF formalism following Datta (1989, 1990), Meir and Wingreen (1992). I will summarize the complete set of equations (section 2), present illustrative examples (Section 3) and conclude with a brief discussion of current research and unanswered questions (Section 4). I have written extensively about the NEGF-Landauer model in the past (Datta 1995, 2005, 2008) and will not repeat any of the detailed derivations or discussions. Neither will I attempt to provide a balanced overview of the vast literature on quantum transport. My purpose is simply to convey our particular viewpoint, namely the bottom-up approach to nanoelectronics, which I believe should be of interest to a broad audience interested in the atomistic description of non-equilibrium phenomena.

## 2. THE NEGF-LANDAUER MODEL

Fig.5 shows a schematic summarizing the basic inputs that define the NEGF-Landauer model. The channel is described by a Hamiltonian $[H_0]$ while the communication between the channel and the contacts is described by the self-energy matrices $[\Sigma_1]$ and $[\Sigma_2]$. The self-energy $[\Sigma_s]$ and the potential [U] describe the interaction with the surroundings and have to be determined self-consistently as we will explain shortly. Each of these quantities is a matrix whose dimension (NxN) depends on the number of basis functions (N) needed to represent the channel. $[H_0]$ and [U] are Hermitian, while $[\Sigma_{1,2,s}]$ have anti-Hermitian components

$$\Gamma_{1,2,s} = i[\Sigma_{1,2,s} - \Sigma_{1,2,s}^+]$$

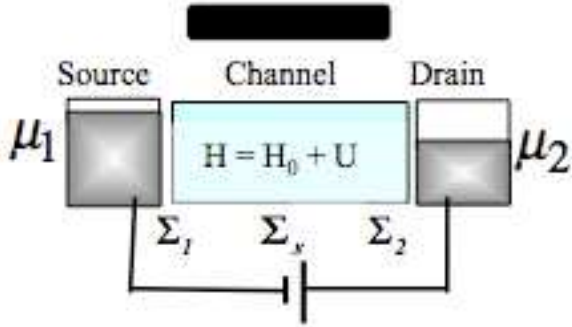

Fig.5: Schematic summarizing the basic inputs that define the NEGF-Landauer model widely used to model nanoscale devices.

All contacts (Fig.5 shows two, labeled source and drain) are assumed to remain in local equilibrium with electrons distributed according to specified Fermi functions

$$f_{1,2}(E) = \frac{1}{1+\exp\left(\dfrac{E-\mu_{1,2}}{k_B T_{1,2}}\right)}$$

Given these inputs, we can calculate any quantity of interest such as the density of states or the electron density or the current using the equations summarized in Section 2.1. But first let me briefly mention a simplified version (Fig.6) that can be obtained from the full NEGF-Landauer model with appropriate approximations as described in Section 2.2. The inputs to this model are the density of states, D(E-U) which floats up or down according to the local potential U, along with escape rates $\gamma_{1,2,s}$ that are simple numbers representing the same physics as the anti-Hermitian part $[\Gamma_{1,2,s}]$ of the self-energy matrices. Despite the simplifications that limit its applicability, this model has the advantage of illustrating much of the essential physics of nanoelectronic devices [Datta 2005, 2008].

For example, in Section 2.2 we obtain the following equation

$$I(E) = \frac{q}{\hbar} \frac{\gamma_1 \gamma_2}{\gamma_1 + \gamma_2} D(E) (f_1(E) - f_2(E)) \qquad \text{(same as Eq.(8))}$$

for the current per unit energy as a special case of the general matrix equations. However, this equation can be obtained from elementary arguments without invoking any advanced concepts, as I do in an undergraduate course on nanoelectronics that I have developed (see chapter 1 of Datta 2005). The point I want to make about Eq.(8) is that it illustrates the basic "force" that drives the flow of current : $f_1(E) - f_2(E)$. Contact 1 tries to fill the states in the channel according to $f_1(E)$, while contact 2 tries to fill them according to $f_2(E)$. As long as $f_1(E) \neq f_2(E)$, one contact keeps pumping in electrons and the other keeps pulling them out leading to current flow. It is easy to see that this current flow is restricted to states with energies close to the electrochemical potentials of the contacts. For energies E that lie far below $\mu_1$ and $\mu_2$, both $f_1(E)$ and $f_2(E)$ are approximately equal to one and there is no steady-state current flow. Although this conclusion appears obvious, it is not necessarily appreciated widely, since many view the electric field as the driving force, which would act on all electrons regardless of their energy. But the real driving force is the difference between the two Fermi functions, which is sharply peaked at energies close to the electrochemical potentials.

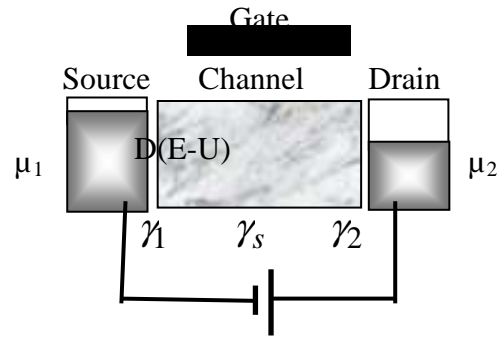

Fig.6: Schematic representing the independent-level model for nanoscale devices which can be viewed as a simple version of the general model of Fig.5 with matrices replaced by ordinary numbers.

Once we recognize the role of $f_1(E) - f_2(E)$ as ***the driving force***, thermoelectric effects are also easily understood. If both contacts have the same electrochemical potential µ, but different temperatures, we have a driving force $f_1(E) - f_2(E)$ that changes sign at E = µ leading to a thermoelectric current whose sign depends on whether the density of states D(E) is increasing or decreasing around E = µ. The molecular Seebeck effect predicted from this argument (Paulsson and Datta 2003) seems to be in good agreement with recent experimental observations (Reddy et al. 2007). This viewpoint also provides a natural explanation for phenomena like the Peltier effect that form the basis for thermoelectric refrigerators (Shakouri 2006). We mentioned earlier that in the Landauer model all the Joule heat is dissipated in the two contacts. But if a conductor has a non-zero density of states only above the electrochemical potentials (Fig.7) then an electron in order to transmit has to first absorb heat from contact 1 thus cooling this contact.

In order for electrons to flow in the direction shown we must have $f_1(E) > f_2(E)$ which requires

$$\frac{E - \mu_1}{T_1} < \frac{E - \mu_2}{T_2}$$

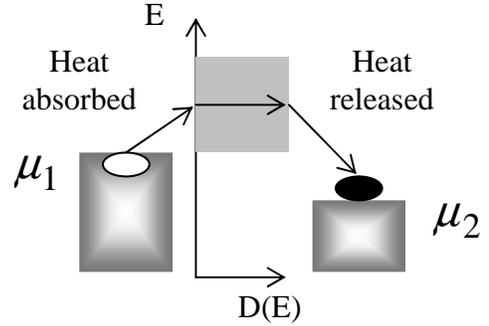

Fig.7: "Peltier effect": If a conductor has a non-zero density of states only above the electrochemical potentials (Fig.7) then an electron in order to transmit has to first absorb heat from contact 1 thus cooling this contact.

Noting that $E - \mu_1$ represents the heat removed from contact 1 and $E - \mu_2$ represents the heat released to contact 2, we recognize this as a statement of the Carnot principle.

What I am trying to illustrate here is the clarity with which many key concepts can be understood within the bottom-up approach, especially if we use the simplified version (Fig.6). However, in this chapter we do not discuss this version any further. Instead we will focus on the full matrix version.

2.1. Summary of equations

A derivation of the basic equations of the NEGF-Landauer method can be found in Datta 2005 both from a one-electron Schrodinger equation (see Chapter 9) and from a second quantized formalism (see Appendix). Here we will simply summarize the equations without derivation. In quantum transport we have a matrix corresponding to each quantity of interest from which the desired information can be extracted. For example, we have a *spectral function* whose diagonal elements give us the local density of states (times $2\pi$), an electron and a hole *correlation function* whose diagonal elements give us the electron and hole density per unit energy (times $2\pi$) and a *current operator* [$I^{op}$] whose trace gives us the current. The following equations allow us to calculate these quantities.

(1) *Spectral function, [A(E)]* is obtained from

$$G(E) = [EI - H_0 - U - \Sigma_1 - \Sigma_2 - \Sigma_s]^{-1} \qquad (2a)$$

$$A(E) = i[G - G^+] \qquad (2b)$$

(2) **Electron and hole correlation functions [$G^n(E)$ and $G^p(E)$]** are obtained from

$$[G^n(E)] = [G\Gamma_1 G^+] f_1 + [G\Gamma_2 G^+] f_2 + [G\Sigma_s^{in} G^+] \qquad (3a)$$

$$[G^p(E)] = [G\Gamma_1 G^+](1 - f_1) + [G\Gamma_2 G^+](1 - f_2) + [G\Sigma_s^{out} G^+] \qquad (3b)$$

It can be shown that $A = G^n + G^p$, as we would expect since the density of states should equal the sum of the electron and hole densities.

(3) **Current operator, $I_i$ at terminal 'i'** per unit energy is obtained from

$$I_i^{op}(E) = \frac{iq}{h}\left([\Gamma_i G^+ - G\Gamma_i] f_i - [\Sigma_i G^n - G^n \Sigma_i^+]\right) \qquad (4a)$$

The charge current per unit energy (to be integrated over all energy for the total current) is obtained from the trace of the current operator:

$$I_i(E) = \frac{q}{h}\left(Trace[\Gamma_i A] f_i - Trace[\Gamma_i G^n]\right) \qquad (4b)$$

while the coherent component of the current can be calculated from the relation

$$I_{coh}(E) = \frac{q}{h} Trace[\Gamma_1 G \Gamma_2 G^+](f_1(E) - f_2(E)) \qquad (4c)$$

where the quantity $\overline{T}_{coh}(E) \equiv Trace[\Gamma_1 G \Gamma_2 G^+]$ is called the "transmission". Eq.(4c) only gives the **coherent** part of the current while Eq.(4b) gives us the full current, the coherent plus the incoherent.

Note that the current operator from Eq.(4a) can be used to calculate other quantities of interest as well. For example, the spin current could be obtained from Trace [$\vec{S} I_i^{op}$] where $\vec{S}$ is an appropriate matrix representing the spin.

Eqs.(2) through (4) involve three quantities [U], [$\Sigma_s$] and [$\Sigma_s^{in}$] that describe the interactions of an individual electron with its surroundings. These quantities are functions of the correlation functions ([$G^n, G^p$]) and have to be calculated *self-consistently*. The actual function we use embodies the physics of the interactions as we will outline below. But let us first neglect these interactions and try to get a physical feeling for Eqs.(2) through (4), by applying them to a particularly simple problem.

2.2. Independent level model

Eqs.(2) through (4) provide a general approach to the problem of quantum transport, with inputs in the form of (NxN) matrices. The Hamiltonian matrix [H] has N eigenstates and a simple approach is to treat each eigenstate separately and add up the currents as if we have N independent levels in parallel. We call this the "independent level model" which would be precisely correct if the self-energy matrices were also diagonalized by the transformation that diagonalizes [H]. This is usually not the case, but the independent level model often provides good insights into the basic physics of nanoscale transport.

Consider a channel with a single energy eigenstate in the energy range of interest. We can use this eigenstate as our basis to write all input parameters as (1x1) matrices or pure numbers:

$$[H] = \varepsilon, \ [\Sigma_1] = -i\gamma_1/2, \ [\Sigma_2] = -i\gamma_2/2, \ [\Gamma_1] = \gamma_1, \ [\Gamma_2] = \gamma_2$$

Neglecting all interactions and setting each of the quantities [U], [$\Sigma_s$] and [$\Sigma_s^{in}$] to zero, we have from Eq.(2) for the

$$\text{Green's function } G = \frac{1}{E - \varepsilon + i(\gamma_1 + \gamma_2)/2} \quad (5a)$$

and the Spectral function $A = \dfrac{\gamma}{(E-\varepsilon)^2 + (\gamma/2)^2}$ where $\gamma \equiv \gamma_1 + \gamma_2$ (5b)

The density of states is equal to $A/2\pi$ showing that the energy level is broadened around the energy level $\varepsilon$. Eq.(5) gives the occupation of this broadened level

$$\text{Electron correlation function } G^n = \frac{\gamma_1 f_1 + \gamma_2 f_2}{(E-\varepsilon)^2 + (\gamma/2)^2} \quad (6a)$$

or the lack of occupation thereof

$$\text{Hole correlation function } G^p = \frac{\gamma_1(1-f_1) + \gamma_2(1-f_2)}{(E-\varepsilon)^2 + (\gamma/2)^2} \quad (6b)$$

The electron and hole density per unit energy are given by $G^n/2\pi$ and $G^p/2\pi$ respectively and as expected, $A = G^n + G^p$.

Finally, the current can be calculated from Eq.(4b) or (4c)

$$I(E) = \frac{q}{h} \frac{\gamma_1 \gamma_2}{(E-\varepsilon)^2 + (\gamma/2)^2} (f_1(E) - f_2(E)) \quad (7)$$

Using Eqs.(7) and (5b) we can write

$$I(E) = \frac{q}{\hbar} \frac{\gamma_1 \gamma_2}{\gamma_1 + \gamma_2} D(E)(f_1(E) - f_2(E)) \quad (8)$$

where $D(E) = A(E)/2\pi$ is the broadened density of states associated with the level.

Now if we superpose the results from N levels we still have exactly the same equation for the current. It is just that D(E) now represents the total density of states rather than just the part associated with a particular level. Indeed one can include a self-consistent potential U into this model simply by letting the density of states float up or down, D(E-U) and this approach (Fig.6) has proved quite successful in providing a simple description of nanoscale transistors [Rahman et al. 2003]. Elastic and inelastic interactions can also be included straightforwardly into this model [Datta 2007]. However, we will not discuss this model further in this chapter. Instead we will focus on the full matrix version.

2.3. Self-consistent potential, [U]

The potential [U] represents the potential that an individual electron feels due to the other electrons and as such we expect it to depend on the electron density or more generally the correlation function $[G^n]$. In semi-empirical theories the Hamiltonian $[H_0]$ often includes the potential under equilibrium conditions, so that [U] itself should account only for the deviation $[\delta G^n]$ from equilibrium. How [U] is related to $[G^n]$ or to $[\delta G^n]$ depends on the approximation used, the simplest being the Hartree approximation which is equivalent to using the Poisson equation or classical electrostatics. More sophisticated theories using many-body perturbation theory or density functional theory will include corrections to account for exchange and correlation. We will not go into this any further, except to note that there are examples of devices whose current-voltage characteristics cannot be described within this approach no matter how sophisticated our choice of "U". These devices seem to require models that go beyond the framework described here (see concluding section).

2.4. Intra-channel interactions: $[\Sigma_s]$ and $[\Sigma_s^{in}]$

As I mentioned earlier, the classic NEGF formalism like much of the pre-mesoscopic literature on transport theory paid no attention to the contacts. Instead it was focused on the quantities $[\Sigma_s]$ and $[\Sigma_s^{in}]$ and provided systematic prescriptions for writing them down using diagrammatic perturbation theoretic treatment to treat interactions [Danielewicz 1984]. In the self-consistent Born approximation (SCBA) we can write for any interaction involving an exchange of energy $\varepsilon$

$$\left[\Sigma_s^{in}(E)\right]_{ij} = D_{ijkl}(\varepsilon)\left[G^n(E-\varepsilon)\right]_{kl} \quad (9a)$$

$$\left[\Sigma_s^{out}(E)\right]_{ij} = D_{lkji}(\varepsilon)\left[G^p(E+\varepsilon)\right]_{kl} \qquad (9b)$$

*where summation over repeated indices is implied.* $[\Sigma_s]$ is obtained as follows: Its anti-Hermitian component is given by $\Gamma_s(E) = \Sigma_s^{in}(E) + \Sigma_s^{out}(E)$, while the Hermitian part is obtained by finding its Hilbert transform.

The "scattering current" is given by (cf.Eq.(4b))

$$I_s(E) = \frac{q}{h}\left(Trace[\Sigma_s^{in}A] - Trace[\Gamma_s G^n]\right) \qquad (10a)$$

$$= \frac{q}{h}\left(Trace[\Sigma_s^{in}G^p] - Trace[\Sigma_s^{out}G^n]\right) \qquad (10b)$$

and it can be shown that $\sum_i I_i(E) + I_s(E)$ is assured to equal zero at all energies, as required for current conservation. Making use of Eqs.(9a,b) we can write Eq.(10) in the form

$$I_s(E) = \frac{q}{h}\sum_{i,j,k,l} D_{ijkl}(\varepsilon)\,G_{kl}^n(E-\varepsilon)\,G_{ji}^p(E) - D_{lkji}(\varepsilon)\,G_{kl}^p(E+\varepsilon)\,G_{ji}^n(E) \qquad (10c)$$

which can be integrated to show that $\int dE\, I_s(E) = 0$, as we would expect since there is no net exchange of electrons with the scatterers. However, $\int dE\, E\, I_s(E) \neq 0$, indicating the possibility of energy exchange. This equation can be understood in semiclassical terms if we assume that the electron and hole matrices are both purely *diagonal*:

$$I_s(E) \to \frac{q}{h}\sum_{i,k} D_{iikk}(\varepsilon)\,G_{kk}^n(E-\varepsilon)\,G_{ii}^p(E) - D_{kkii}(\varepsilon)\,G_{kk}^p(E+\varepsilon)\,G_{ii}^n(E)$$

This is essentially the standard scattering term in the Boltzmann equation if we associate the D tensor with the scattering probabilities: $D_{iikk}(\varepsilon) \to S_{ik}(\varepsilon)$. We know from the Boltzmann treatment that if the entity (like phonons) with which the electrons interact is in equilibrium with temperature $T_s$, then in order to comply with the laws of thermodynamics, we must have $S_{ik}(\varepsilon) = S_{ki}(-\varepsilon)\exp(-\varepsilon/k_B T_s)$. The corresponding relation in quantum transport

$$D_{ijkl}(\varepsilon) = D_{lkji}(-\varepsilon)\exp(-\varepsilon/k_B T_s) \qquad (11)$$

is more subtle and less appreciated. Note, however, that neither the semiclassical nor the quantum restriction is operative, if the interacting entity is not in equilibrium.

If we assume *elastic interactions ($\varepsilon = 0$), along with the equilibrium condition (Eq.(11))*, then we can write

$$\left[\Sigma_s^{in}(E)\right]_{ij} = D_{ijkl}\left[G^n(E)\right]_{kl} \quad \text{and} \quad \left[\Sigma_s^{out}(E)\right]_{ij} = D_{ijkl}\left[G^p(E)\right]_{kl} \quad (12)$$

so that $\quad \left[\Gamma_s(E)\right]_{ij} = D_{ijkl}\left[A(E)\right]_{kl}$

and [$\Sigma_s$] can be related directly to [G]:

$$\left[\Sigma_s\right]_{ij} = D_{ijkl}\left[G\right]_{kl} \quad (13)$$

This simplifies the calculation by decoupling Eqs.(2) from (3) but it is important to note that Eqs.(12) and (13) are valid only fro elastic interactions with scatterers that are in equilibrium.

As mentioned above, the NEGF formalism provides clear prescriptions for calculating the tensor [[D]] starting from any given microscopic interaction Hamiltonian. Alternatively, we have advocated a phenomenological approach whereby specific choices of the form of the tensor [[D]] give rise selectively to phase, momentum or spin relaxation and their magnitudes can be adjusted to obtain desired relaxation lengths for these quantities as obtained from experiment For example, the following choice (Golizadeh-Mojarad and Datta, 2007).

$$D_{ijkl} = d_p\,\delta_{ik}\delta_{jl} \quad (14a)$$

$d_p$ being a constant, leads to pure phase relaxation. This is equivalent to writing [$\Sigma_s$] and [$\Sigma_s^{in}$] as a constant times [$G$] and [$G^n$] respectively:

$$\left[\Sigma_s\right]_{ij} = d_p\left[G\right]_{ij} \quad \text{and} \quad \left[\Sigma_s^{in}\right]_{ij} = d_p\left[G^n\right]_{ij} \quad (14b)$$

I will present a concrete example showing that this choice of the tensor [[D]] indeed relaxes phase without relaxing momentum. But one can see the reason intuitively by noting that the SCBA (Eq.(9)) effectively takes electrons out of the channel and feeds them back with a randomized phase similar in concept to the Buttiker probes widely used in mesoscopic physics [Datta 1989, Hershfield 1991]. A constant multiplier as shown in Eq.(23b) suggests that the electrons are fed back *while preserving the initial correlation function exactly*. We thus expect no property of the electrons to be relaxed except for phase.

Another choice $\quad D_{ijkl} = d_m\,\delta_{ij}\,\delta_{ik}\,\delta_{jl} \quad (15a)$

that we will illustrate is equivalent to writing

$$[\Sigma_s]_{ij} = d_m \, \delta_{ij} \, [G]_{ij} \quad \text{and} \quad \left[\Sigma_s^{in}\right]_{ij} = d_p \delta_{ij} \left[G^n\right]_{ij} \quad (15b)$$

Unlike the phase relaxing choice (Eqs.(23)), this choice feeds back only the diagonal elements. In a real space representation this leads to momentum relaxation in addition to phase relaxation, as we will see in Section 3.

A choice that leads to pure spin relaxation is $\quad D_{abcd} = d_s \, \vec{\sigma}_{ac} \bullet \vec{\sigma}_{db} \quad (16a)$

where we have used a separate set of indices (a,b,c,d instead of i,j,k,l) to indicate that these are spin indices. The tensor has the same form as that for pure phase relaxing interactions (Eq.(23)) as far as the indices other than spin are concerned. Here $\vec{\sigma}$ denotes the Pauli spin matrices and Eq.(25a) is equivalent to writing

$$[\Sigma_s] = d_s \, ([\sigma_x][G][\sigma_x] + [\sigma_y][G][\sigma_y] + [\sigma_z][G][\sigma_z])$$

and $\quad \left[\Sigma_s^{in}\right] = d_s \, ([\sigma_x][G^n][\sigma_x] + [\sigma_y][G^n][\sigma_y] + [\sigma_z][G^n][\sigma_z]) \quad (16b)$

It is straightforward to show that $Trace\left[\Sigma_s^{in} \vec{\sigma}\right] = -\, d_s \, Trace \, [G^n \vec{\sigma}]$, indicating that this choice for the tensor [[D]] feeds back a spin equal to $-d_s$ times the original spin, thus leading to spin relaxation.

In the next section we present a few examples to give the reader a flavor of how these equations are applied. More examples, especially those involving spin are discussed in another chapter in this volume [Golizadeh-Mojarad and Datta].

3. A FEW EXAMPLES

3.1. Single-moded channel

Consider first a one-dimensional single-band tight-binding model with a nearest neighbor Hamiltonian of the form

$$\begin{bmatrix} \varepsilon & -t & 0 & 0 & \cdots \\ -t & \varepsilon & -t & 0 & \cdots \\ 0 & -t & \varepsilon & -t & 0 & \cdots \end{bmatrix} \quad (17)$$

which can be represented schematically as shown in Fig.8. In principle, the Hamiltonian should also include the potential due to any external voltages applied to the electrodes, but for our examples we will neglect it assuming it to be small. We will also ignore the self-consistent potential [U].

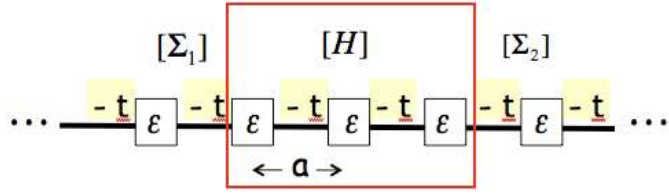

Fig.8: One-dimensional single-band tight-binding model with site energy $\varepsilon$ and nearest neighbor overlap "-t" having a dispersion relation of the form E = $\varepsilon$ - 2t cos ka, a being the nearest neighbor distance.

Let us treat just one site as the channel $[H_0] = \varepsilon$ and the rest of the semi-infinite wire on either side as self-energies that are given by (Caroli et al. 1972)

$$[\Sigma_1] = -te^{ika} \text{ and } [\Sigma_2] = -te^{ika}$$

so that $[\Gamma_1] = 2t \sin ka$ and $[\Gamma_2] = 2t \sin ka$

where "ka" is related to the energy by the dispersion relation E = $\varepsilon$ - 2t cos ka.

From Eq.(2), $\quad [G] = \dfrac{1}{E - \varepsilon + 2te^{ika}} = \dfrac{-i}{2t \sin ka}$

From Eq.(4), $\quad I(E) = (q/h)(f_1(E) - f_2(E))$ (18)

as long as $\varepsilon$-2t < E < $\varepsilon$+2t. Outside this energy range, "ka" is imaginary, making $[\Sigma_1]$ and $[\Sigma_2]$ purely real and hence $[\Gamma_1] = [\Gamma_2] = 0$.

From Eq.(18) we obtain for the total current

$$I = (q/h) \int dE \, (f_1(E) - f_2(E)) = (q/h)(\mu_1 - \mu_2)$$

Since $\mu_1 - \mu_2 = qV$ this shows that a one-dimensional ballistic wire has a conductance equal to the quantum of conductance: $I/V = q^2/h$.

Note that the single-band tight-binding Hamiltonian in Eq.(17) can alternatively be viewed as a discrete version of a one-dimensional effective mass Hamiltonian of the form $-\dfrac{\hbar^2}{2m} \dfrac{\partial^2}{\partial x^2}$, if we set $t = \hbar^2/2ma^2$, $\varepsilon$=2t. Any potential U(x) can be included in Eq.(17) by adding $U(x = x_i)$ to each diagonal element (i,i). The continuum version has a dispersion relation $E = \hbar^2 k^2/2m$ while the discrete version has a dispersion relation $E=2t(1-\cos ka)$. The two agree reasonably well for ka < $\pi/3$, with energies in the range 0 < E < t.

## 3.2. Conductance quantization

Fig.9 shows the transmission versus energy calculated for a rectangular conductor of width 102 nm using the model described below. Note the discrete integer steps in the transmission as the energy increases and new subbands or transverse modes come into play. The discrete integer values for the transmission lead to low bias conductance values that are approximate integer multiples of the conductance quantum. This quantization of the conductance in multi-moded wires, first observed experimentally in 1988 (van Wees et al. 1988, Wharam et al. 1988) serves as a good benchmark for any theory of quantum transport.

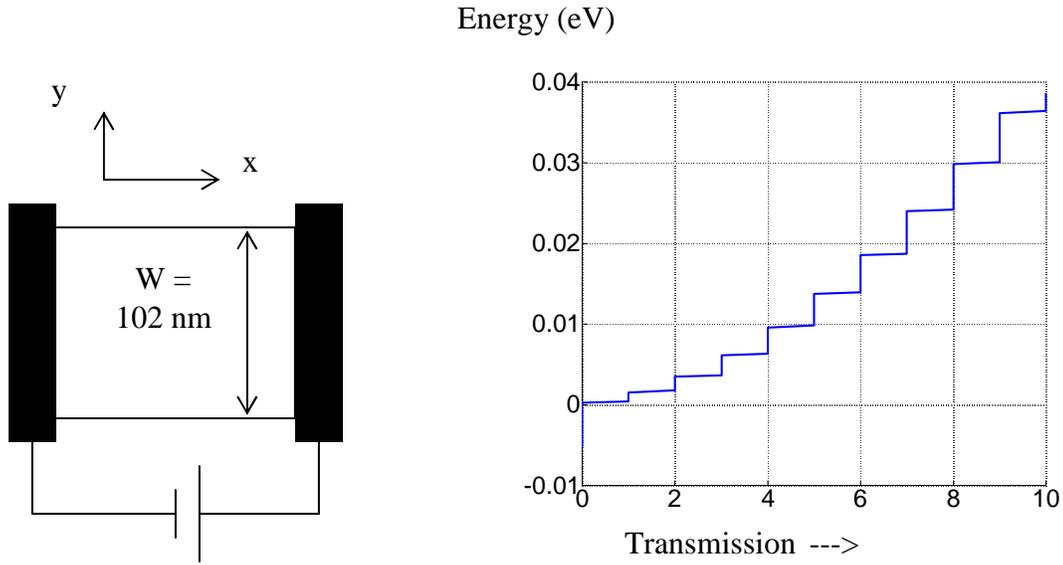

Fig.9: Transmission versus energy for a rectangular conductor of width 102 nm modeled with a single-band tight-binding model with $t = \hbar^2/2ma^2 \approx 0.04$ eV, $\varepsilon = 4t$, m = 0.25*free electron mass, a = 2 nm.

### 3.2.1. Model details

The rectangular conductor is modeled with a single-band tight-binding model with $t = \hbar^2/2ma^2 \approx 0.04$ eV, $\varepsilon = 4t$, m = 0.25*free electron mass and a = 2 nm (Fig.10). Conceptually we can lump each column of the square lattice into a single matrix $\alpha$, which is essentially the one-dimensional Hamiltonian from the last section (Eq.(17)). Neighboring columns are coupled by a matrix $\beta$ to the left and $\beta^+$ to the right. In this example, $\beta = \beta^+ = -t[I]$, [I] being the identity matrix, but in general $\beta$ need not equal $\beta^+$.

The overall Hamiltonian is written as

$$\begin{bmatrix} \alpha & \beta & 0 & 0 & \cdots \\ \beta^+ & \alpha & \beta & 0 & \cdots \\ 0 & \beta^+ & \alpha & \beta & 0 & \cdots \end{bmatrix} \qquad (19)$$

The contact self-energies are given by $\Sigma_1 = \beta g_1 \beta^+$ and $\Sigma_2 = \beta^+ g_2 \beta$ where $g_1$ and $g_2$ are the surface Green's functions for the left and right contacts respectively (they are the same in this example, but need not be in general). These surface Green's functions can be obtained by solving the matrix quadratic equations

$$[g_1]^{-1} = \alpha - \beta g_1 \beta^+ \text{ and } [g_2]^{-1} = \alpha - \beta^+ g_2 \beta \qquad (20)$$

These can be solved iteratively in a straightforward manner but this can be time-consuming for wide conductors and special algorithms may be desirable. If the matrices $\alpha$ and $\beta$ can be simultaneously diagonalized then a faster approach is to use this diagonal basis to write down the solutions to Eq.(20) and then transform back. In this basis the multi-moded wire decouples into separate single-moded wires. However, this simple decoupling is not always possible since the same unitary transformation may not diagonalize both $\alpha$ and $\beta$.

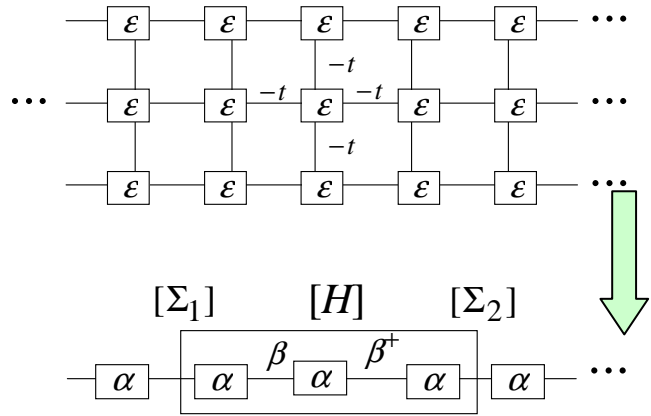

Fig.10: Single-band tight-binding model on a square lattice with site energy $\varepsilon$ and nearest neighbor overlap "-t" having a dispersion relation of the form E = $\varepsilon - 2t \cos k_x a - 2t \cos k_y a$, a being the nearest neighbor distance. Conceptually we can lump each column into a single matrix $\alpha$, with neighboring columns coupled by a matrix $\beta$ to the left and $\beta^+$ to the right.

3.3. Ballistic Hall effect

Fig.11 shows another interesting result, namely the Hall resistance normalized to the resistance quantum ($h/e^2$) as a function of the magnetic field (applied along the z-direction) calculated for a rectangular conductor of width W = 102 nm. Note the plateaus in the Hall resistance equal to the inverse of integers 2,3,4 etc. representing the quantum Hall effect. This calculation is done using essentially the same model as in the last example, but there are two additional points that need clarification.

The first point is that the magnetic field $\vec{B} = B\hat{z}$ enters the Hamiltonian through the phase of the nearest neighbor coupling elements as shown in Fig.12. The second point is the concept of a *local electrochemical potential* that we have used to obtain the Hall voltage. Our calculations are done at a single electron energy E and at this energy we assume the Fermi functions $f_1(E)$ and $f_2(E)$ to equal one and zero respectively. At all points "i" within the channel, the occupation lies between 0 and 1, and it is this occupation that we call the local electrochemical potential and estimate it from the ratio of the local electron density to the local density of states [McLennan et al.1991]:

$$\mu(i) = G^n(i,i)/A(i,i) \qquad (21)$$

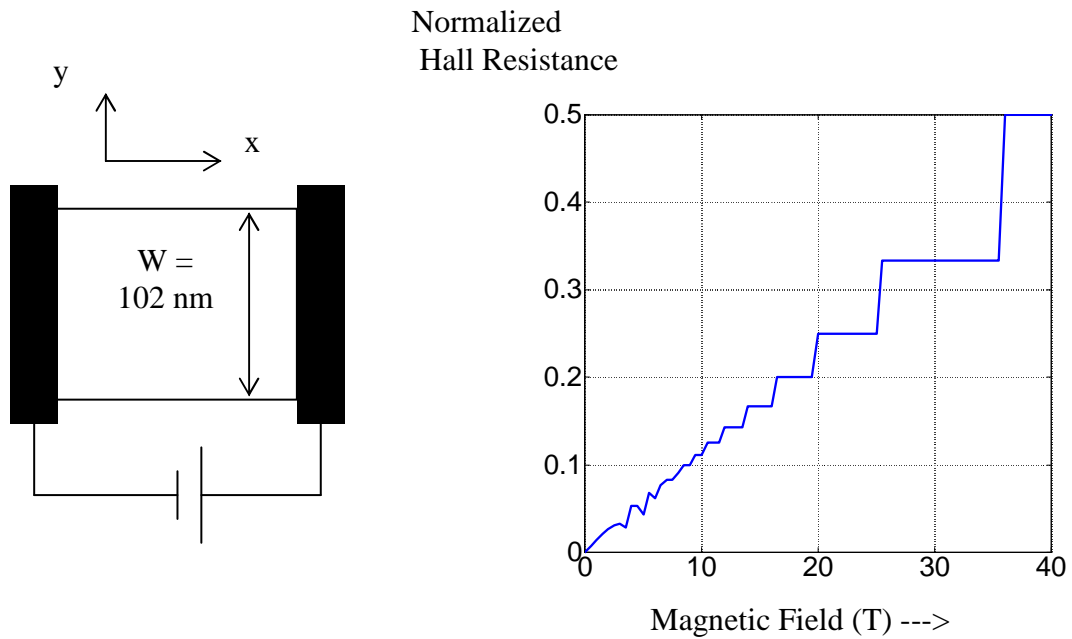

Fig.11: Hall resistance ( = Hall voltage / current) normalized to the resistance quantum ($h/e^2$) versus magnetic field (applied along the z-direction) calculated for a rectangular conductor of width W = 102 nm. Note the plateaus in the Hall resistance equal to the inverse of integers 2,3,4 etc. representing the quantum Hall effect. Electron energy = t ~ 0.04 eV.

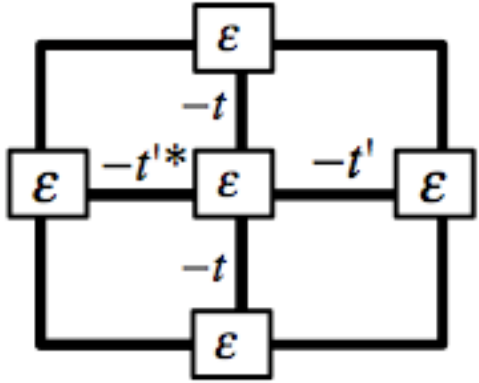

Fig.12. The magnetic field $\vec{B} = B\hat{z}$ represented through a vector potential $\vec{A} = -By\hat{x}$, appears in the single-band tight-binding model in the phase of the coupling elements along x: $t' = t\exp(+iBya)$.

Fig.13 shows a plot of this local electrochemical potential μ across the width of the conductor. At zero magnetic field, μ is constant (= 0.5) and develops a slope as the field is increased. The oscillations arise from coherent interference effects that usually get washed out when we sum over energies or include phase relaxation processes. Here we have estimated the Hall voltage simply by looking at the difference between μ at the two edges of the conductor and the Hall resistance in Fig.11 is obtained by dividing this transverse Hall voltage by the current.

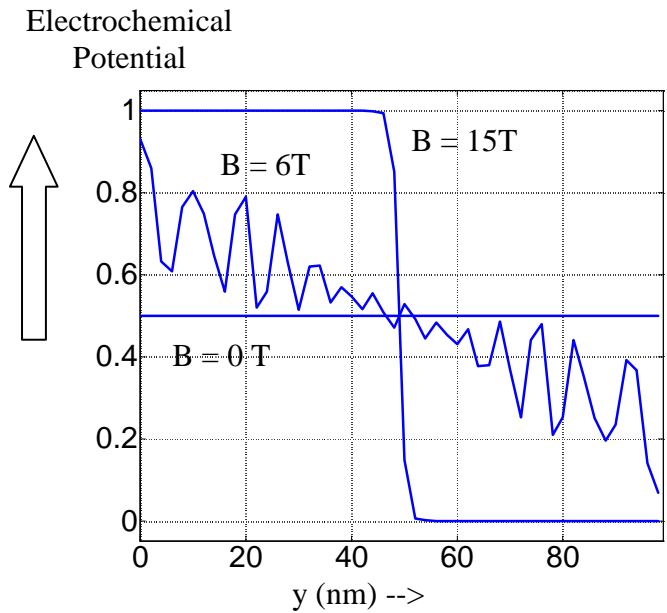

Fig.13. Profile of the local electrochemical potential (defined in Eq.(21)) across the width of the conductor at three different values of the magnetic field. Electron energy = t ~ 0.04 eV.

### 3.4. "Potential" drop across a single-moded channel

An instructive example to look at is the variation of the electrochemical potential (defined by Eq.(21)) across a scatterer in a single-moded wire modeled with a tight-binding model as described in Section 3.1. As expected, the potential drops sharply across the scatterer (Fig.14), but a purely coherent calculation usually yields oscillations arising from interference effects (see Fig.14a). Such oscillations are usually strongly muted if not washed away in room temperature measurements, because of strong phase relaxation. Much of the phase relaxation arises from electron-electron interactions, which to first order do not give rise to any momentum relaxation. Such processes could be included by including an interaction self-energy of the form shown in Eq.(23) and indeed it suppresses the oscillations (Fig.14b). The momentum relaxing interaction shown in Eq.(24) also suppresses oscillations, but it leads to an additional slope across the structure (Fig.14c) as we would expect for a distributed resistance.

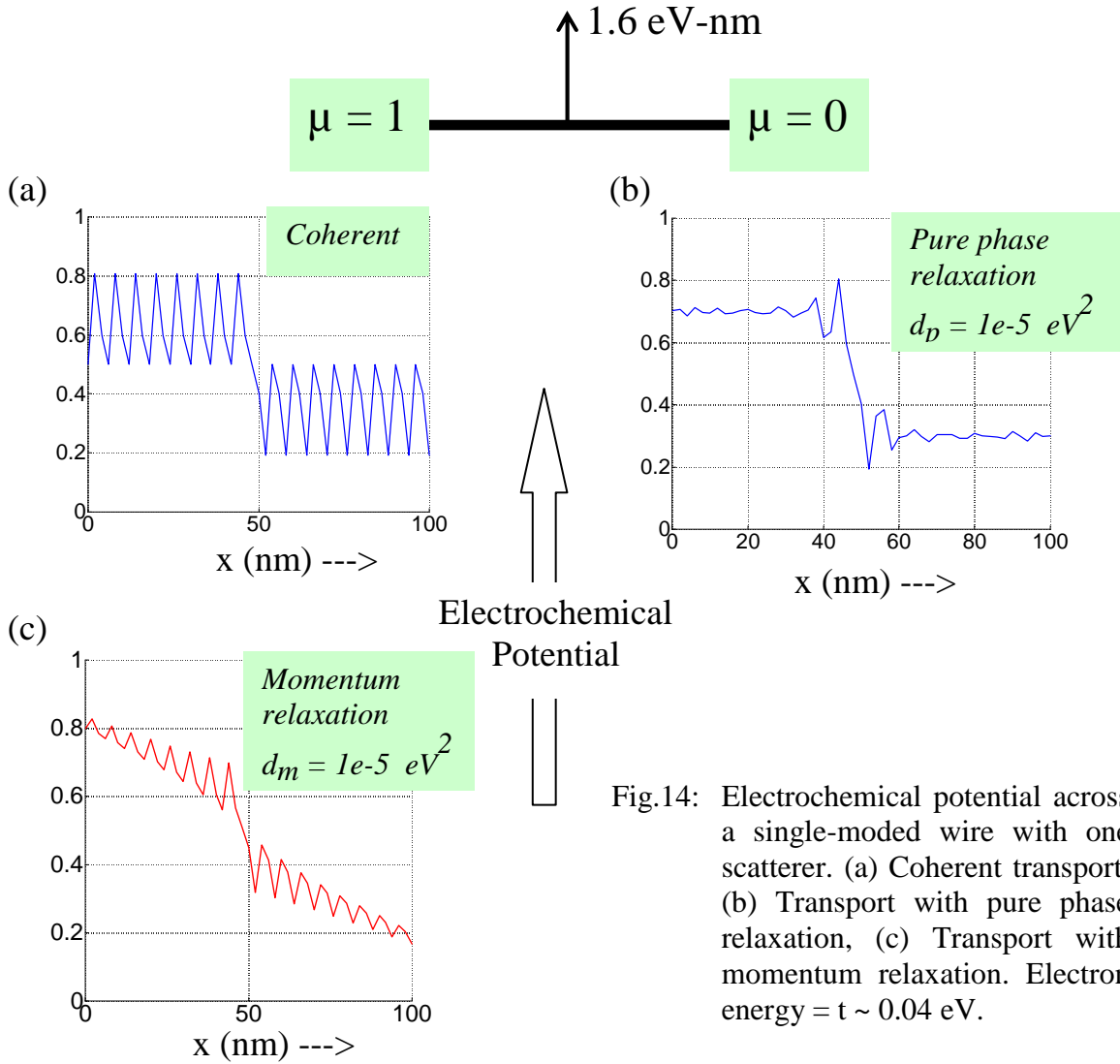

Fig.14: Electrochemical potential across a single-moded wire with one scatterer. (a) Coherent transport, (b) Transport with pure phase relaxation, (c) Transport with momentum relaxation. Electron energy = t ~ 0.04 eV.

3.5. "Potential" drop across a single-moded channel including spin

Another interesting example is the variation of the electrochemical potential for the up-spin and down-spin channels across a single-moded wire connected to anti-parallel ferromagnetic contacts assumed to have a coupling to the majority spin that is (1+P)/(1-P) times the coupling to the minority spin (P=0.95). The up-spin channel is strongly coupled to the contact with µ = 1 and weakly coupled to the contact with µ = 0, with the roles reversed for the down-spin channel. Consequently the electrochemical potential for the up-spin channel is closer to 1 while that for the down-spin channel is closer to 0 (Fig.15a). The difference is reduced when we introduce a little spin-orbit coupling (Fig.15b), but with strong spin-orbit coupling the potential actually oscillates back and forth. This oscillation is the basis for many "spin transistor" proposals (for a recent review see Bandyopadhyay and Cahay 2008), but it should be noted that we are assuming a contact efficiency (95%) that is considerably better than the best currently available. Also our calculations include pure phase relaxation ($d_p = 1e-5\,eV^2$) to account for electron-electron interactions. These processes reduce any oscillations due to multiple spin-independent reflections. Finally Fig.15d shows the effect of spin relaxing processes (Eq.(25)) in equalizing up-spin and down-spin electrochemical potentials.

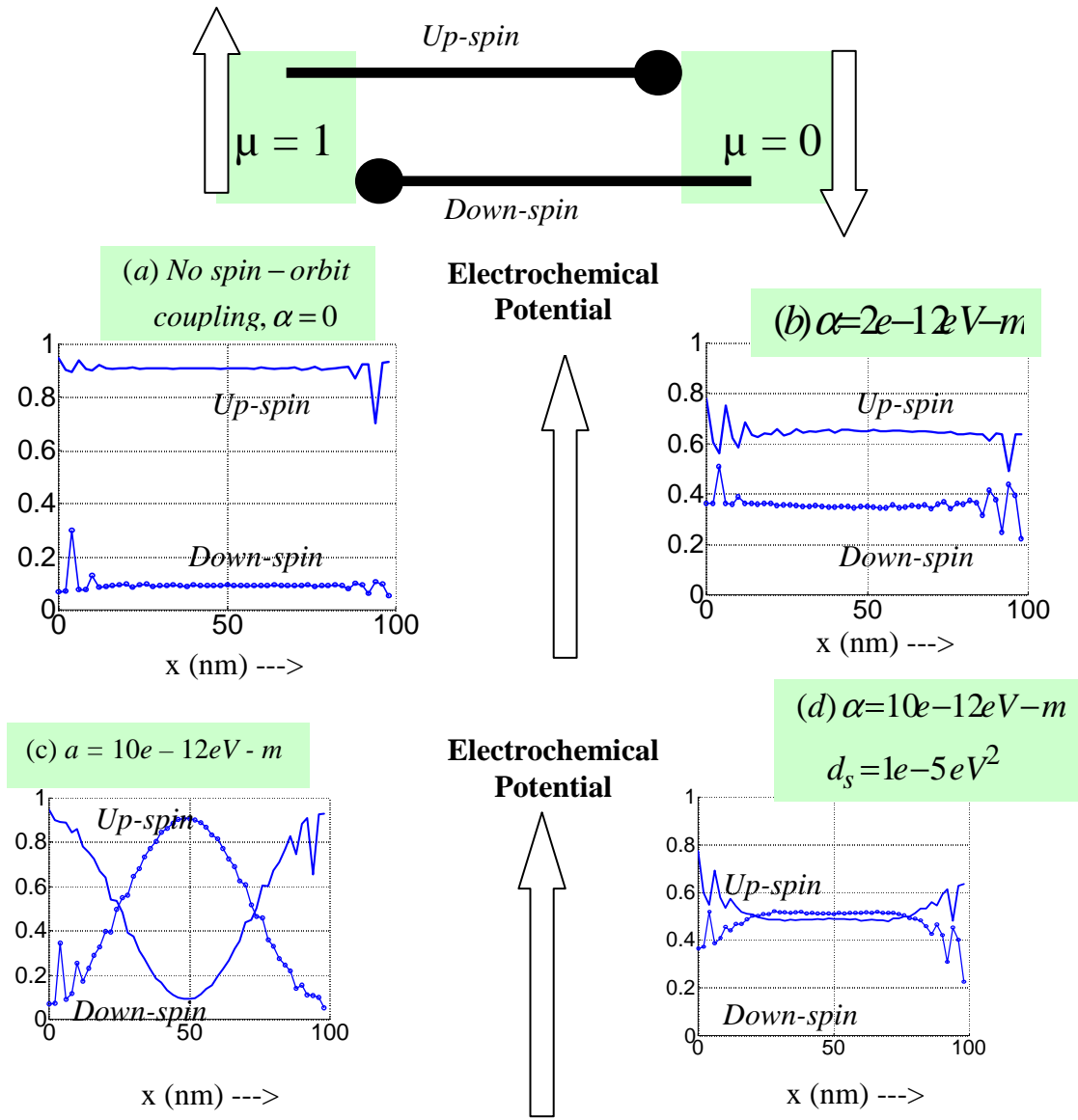

Fig.15. Electrochemical potential for the up-spin and down-spin channels across a single-moded wire connected to anti-parallel ferromagnetic contacts assumed to have a coupling to the majority spin that is (1+P)/(1-P) times the coupling to the minority spin (P=0.95): Ballistic conductor with (a) weak spin-orbit coupling, (b) weak spin-orbit coupling, (c) strong spin-orbit coupling and finally (d) a conductor with spin relaxation in addition to strong spin-orbit coupling. All calculations include pure phase relaxation ($d_p = 1e-5\,eV^2$), which reduce oscillations due to multiple spin-independent reflections.

3.5.1. Model details

A brief explanation of how we include spin-orbit coupling into the single-band tight-binding or effective mass equation described in Section 3.2.1. Conceptually each "grid point" effectively becomes two grid points when we include spin explicitly and so the site energy becomes $\varepsilon[I]$, [I] being a (2x2) identity matrix and the nearest neighbor coupling elements become $-t[I]$. Spin-orbit coupling modifies these coupling elements as shown in Fig.16 which add to the usual $-t[I]$ (not shown). It is straightforward to show that this Hamiltonian leads to a dispersion relation

$$E = (\varepsilon - 2t\cos ka)[I] + \frac{\alpha}{a}([\sigma_x]\sin k_y a - [\sigma_y]\sin k_x a) \qquad (22a)$$

which for small "ka" reduces to the effective mass-Rashba Hamiltonian [Bychkov and Rashba 1984]:

$$E = \frac{\hbar^2 k^2}{2m}[I] + \alpha([\sigma_x]k_y - [\sigma_y]k_x) \qquad (22b)$$

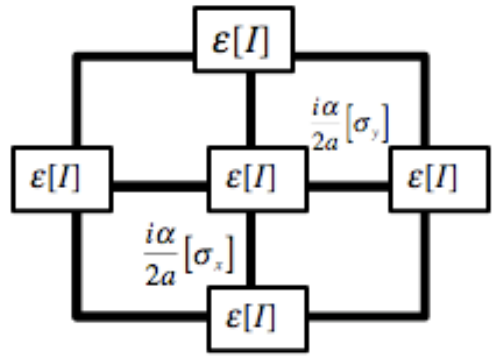

Fig.16: Rashba spin-orbit Hamiltonian on a discrete lattice.

## 4. CONCLUDING REMARKS

A central point that distinguishes our approach based on the NEGF-Landauer method is the explicit acknowledgement of the important role played by the ***contacts***, a role that was highlighted by the rise of mesoscopic physics in the late 1980's. Indeed we are arguing for a bottom-up approach to electronic devices that starts from the coherent or Landauer limit where there is a clear separation between the role of the channel and the contact. The channel is governed by purely dynamical forces, while the contacts are held in local equilibrium by entropic forces. This separation provides a conceptual clarity that makes it very attractive pedagogically, not just for ballistic transport but for all non-equilibrium processes in general. Dynamic and entropic processes are generally intertwined and even the channel experiences entropic forces like the contacts, as long as it has degrees of freedom such as phonons that can be excited. One could say that contacts are not just the physical ones at the ends of the conductor described by $[\Sigma_{1,2}]$. Abstract contacts of all kinds described by $[\Sigma_s]$ are distributed throughout the channel.

Usually all these contacts are assumed to be held in equilibrium by entropic forces. In practice, it is not uncommon for contacts, especially "nanocontacts", to be driven out-of-equilibrium. This is true of physical contacts made to nanotransistor channels, as well as abstract contacts like the non-itinerant electrons in nanomagnets driven by spin-torque forces or the nuclear spins in semiconductors driven by the Overhauser effect. Such out-of-equilibrium "contacts" can be included straightforwardly into the model we have described by coupling the NEGF-Landauer model to a dynamic equation describing the out-of-equilibrium entity, like the Bloch equation for isolated spins or the Landau-Lifshitz-Gilbert (LLG) equation for nanomagnets [see for example, Salahuddin and Datta 2006 and references therein].

The real conceptual problem arises when we allow for the possibility of correlations or entanglement. This can be understood from a simple example. Consider a channel with just two spin-degenerate levels (Fig.17) biased such that contact 1 wants to fill both levels and contact 2 wants to empty them. If both contacts are equally coupled, we would expect each level to be half-filled:

$f_{up} = 0.5$ and $f_{dn} = 0.5$

This is exactly what we would get if we applied the methods discussed in this chapter to this simple problem.

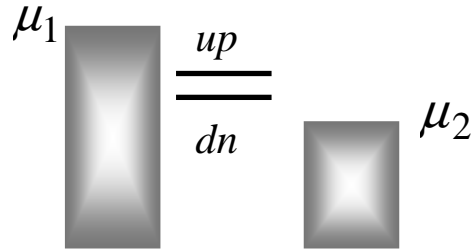

Fig.17. A channel with two spin-degenerate levels biased so that contact 1 wants to fill each level and contact 2 wants to empty them both. Assuming both contacts to be equally coupled to the channel, we would expect each state to be half-filled at steady state.

Now if we ask for the probability that the up-spin level is filled and the down-spin level is empty P(10) we can write it as $f_{up}(1 - f_{dn})$. We can write the probabilities of all four possibilities as

$$P(00) = (1 - f_{up})(1 - f_{dn}) \quad , \quad P(01) = (1 - f_{up}) f_{dn}$$

$$P(10) = f_{up}(1 - f_{dn}) \quad , \quad P(11) = f_{up} f_{dn} \quad (23)$$

In this case this yields P(00) = P(01) = P(10) = P(11) = 1/4.

However, if the electrons are strongly interacting then the energy cost of occupying both levels can be so high that the state (11) has zero probability. Indeed it can be shown that under these conditions P(00) = P(01) = P(10) = 1/3 and P(11) = 0. The point I want to make is that there is no possible choice of $f_{up}$ and $f_{dn}$ that when inserted into Eq.(23) will lead to this result! Since P(11) = 0 we must have either $f_{up}$ or $f_{dn}$ equal to zero, so that P(01) or P(10) would have to be zero. There is no way to obtain non-zero values for both P(01) and P(10), while making P(11) equal zero.

This is an example of a "strong correlation" where the dynamics of individual electrons is so correlated by their interaction that it is inaccurate to view each electron as moving in a mean field due to the other electrons. This "Coulomb blockade" regime has been widely discussed [see for example, Likharev 1999, Beenakker 1991, Braun et al. 2004, Braig and Brouwer 2005] and it can have an important effect on the current-voltage characteristics of molecular scale conductors [Muralidharan 2006] if the single electron charging energy is well in excess of the broadening as well as the thermal energy.

My purpose, however, is not to talk about Coulomb blockade in particular. I use this example simply to illustrate the meaning of correlation and the conceptual issues it raises. One can no longer "disentangle" different electrons. Instead one has to solve a multi-electron problem and a complete transport theory is not yet available in such a multiparticle framework. This is true not just for correlated electrons, but for electrons correlated to other entities such as nuclear spins as well. Any interaction generates correlations, but the standard approach in transport theory is to neglect them following the example of Boltzmann who ignored them through his assumption of "molecular chaos" or "Stohsslansatz", leading to the increase of entropy characteristic of irreversible processes. . Exactly how such multiparticle correlations are destroyed will hopefully become clearer as more delicate experiments are conducted leading to the next level of understanding in transport theory involving "correlated contacts". In the meantime there are many electronic devices for switching, energy conversion and sensing that can be analyzed and designed using the conceptual framework that has emerged in the last twenty years, starting from the Boltzmann (semiclassical dynamics) or the NEGF description (quantum dynamics) appropriate for weak interactions, but extending them to include the contacts. Indeed the distinguishing feature of this framework is the explicit acknowledgement of contacts, leading naturally to a bottom-up approach, which we believe can be very powerful both for teaching and research.

This work was supported by the NSF-sponsored Network for Computational Nanotechnology (NCN) and the Intel foundation.

REFERENCES

This is a very limited set of references directly related to the viewpoint and discussion in this chapter. It is by no means comprehensive or even representative of the vast literature on quantum transport.

Bandyopadhyay, S. and Cahay, M. (2008) Introduction to Spintronics, Taylor & Francis.
Beenakker, C.W.J. (1991) Phys. Rev., **B44**, 1646.
Braig, S. and Brouwer, P.W. (2005) Phys. Rev., **B71**, 195324.
Braun, M., Koenig, J. and Martinek, J. (2004) Phys. Rev., **B70**, 195345.
Bychkov, Y.A. and Rashba, E.I. (1984) J.Phys.C **17**, 6039.
Caroli, C., Combescot, R., Nozieres, P. and Saint-James, D. (1972) J.Phys.C: Solid State Phys. **5**, 21.
Danielewicz, P. (1984) Ann.Phys., NY, **152**, 239.
Datta, S. (1989) Phys. Rev., **B40**, 5830.
Datta, S. (1990) Journal of Physics: Condensed Matter, **2**, 8023.
Datta, S. (1995) Electronic transport in mesoscopic systems, Cambridge University Press.
Datta, S. (2005) Quantum Transport: Atom to Transistor, Cambridge University Press.
Datta, S. (2006) Concepts of Quantum Transport, a series of video lectures, http://www.nanohub.org/courses/cqt
Datta, S. (2008) "Nanodevices and Maxwell's demon", Lecture Notes in Nanoscale Science and Technology, Vol. 2, Nanoscale Phenomena: Basic Science to Device Applications, Eds. Z.K. Tang and P.Sheng, Springer, arXiv:condmat0704.1623.
Feynman, R.P. (1972) Statistical Mechanics. Frontiers in Physics. Addison-Wesley.
Hershfield S. (1991) Phys. Rev. **B43**, 11586.
Kadanoff and Baym (1962) Quantum Statistical Mechanics. Frontiers in Physics Lecture Notes. Benjamin/Cummings.
Keldysh, L.V. (1965) Sov.Phys.JETP, **20**, 1018.
Koswatta, S. O., Hasan, S., Lundstrom. M. S., Anantram. M.P., Nikonov. Dmitri P. (2007) IEEE Trans. Electron Dev., **54**, 2339.
Likharev, K. (1999) Proc. IEEE, **87**, 606.
Martin, P.C. and Schwinger, J. (1959) Phys.Rev. **115**, 1342.
McLennan, M.J., Lee, Y., and Datta, S. (1991) Phys. Rev. B, **43**, 13846.
McQuarrie, D.A. (1976) Statistical Mechanics, Harper and Row.
Meir, Y. and Wingreen, N.S. (1992) Phys. Rev. Lett. **68**, 2512.
Muralidharan, B., Ghosh, A.W., and Datta, S. (2006) Phys. Rev. **B73**, 155410.
Paulsson, M. and Datta, S. (2003) Phys. Rev., **B67**, 241403(R).
Rahman, A., Guo, J., Datta, S., and Lundstrom, M.S. (2003) IEEE Trans. Electron Dev. 50, 1853.
Reddy, P., Jang, S.Y., Segalman, R., and Majumdar A. (2007) Science **315,**1568.
Ren, Z., Venugopal, R., Goasguen, S., Datta, S. and Lundstrom, M. S. (2003) IEEE Trans. Electron. Dev., **50**, 1914.
Salahuddin, S. and Datta, S. (2006) Appl. Phys. Lett., **89**, 153504.
Shakouri, Ali. (2006) Proc. IEEE, **94**, 1613.
van Wees, B.J., van Houten, H., Beenakker, C.W.J., Williamson, J.G., Kouwenhoven, L.P., van der Marel, D., Foxon, C.T. (1988) Phys.Rev.Lett. **60**, 848.
Wharam, D.A., Thornton, T.J., Newbury, R., Pepper, M., Ahmed, H., Frost, J.E.F., Hasko, D.G., Peacock, D.C., Ritchie, D.A. and Jones, G.A.C. (1988) J.Phys.C. **21**, L209.